\documentstyle[aps,prl,epsf]{revtex}

\begin{document}
\twocolumn[{\hsize\textwidth\columnwidth\hsize\csname
@twocolumnfalse\endcsname
\draft
\title{
Stochastic Multiresonance
}
\author{
J. M. G. Vilar
and J. M. Rub\'{\i}
}
\address{
Departament de F\'{\i}sica Fonamental, Facultat de
F\'{\i}sica, Universitat de Barcelona, Diagonal 647,
E-08028 Barcelona, Spain 
\date{\today}
}
\maketitle
\begin{abstract}

We present a class of systems for which the signal-to-noise ratio as
a function of the noise level may display a multiplicity of maxima.
This phenomenon, referred to as stochastic multiresonance, indicates
the possibility that periodic signals may be enhanced at multiple
values of the noise level, instead of at a single value which has
occurred in systems considered up to now in the framework of
stochastic resonance.

\end{abstract}
}]
\narrowtext

In recent years, the phenomenon of stochastic resonance (SR) has
been subject of intense activity, to the extent that many examples
have been found in different scientific areas
\cite{Benzi,tri,Ma1,Ma2,neu1,JSP,neu2,%
Moss,Wies,Wiese,array,thre,grifo,phi4,mio}.  The phenomenon has been
characterized by the appearance of a maximum in the output
signal-to-noise ratio (SNR) at a non-zero noise level.  In this
sense, noise plays a constructive role since an optimized amount is
responsible for the enhancement of the response of the system to a
periodic signal, which otherwise would be manifested with more
difficulty.  In spite of the efforts devoted to its understanding,
there is an aspect which has not been considered up to now.  Under
some circumstances, the SNR may display a multiplicity of maxima and
hence, there is a set of values of the noise level at which the
response of the system is enhanced.  In this Letter we address
precisely this possibility for the appearance of those maxima, then
we show that the concept of SR is more general than the one we
already know.  In this regard, we have found a class of systems
whose output SNR is a non-trivial periodic function of the logarithm
of the noise level.  Similarly, our analysis also reveals that the
SNR as a function of the noise level may exhibit any number of
maxima.

Consider systems with only one relevant degree of freedom whose
dynamics is described by the following Langevin equation
\begin{equation} \label{model}
{dx \over dt} = -f(x,t)x +\sqrt{D}\xi(t) \;\; ,
\end{equation}
where $f(x,t)$ is a given function, $\xi(t)$ is Gaussian white noise
with zero mean and second moment
$\left<\xi(t)\xi(t^\prime)\right>=\delta(t-t^\prime)$, and $D$ is a
constant defining the noise level.  Here the input signal enters the
system through $f(x,t)$, and we will assume it to be periodic in
time with frequency $\omega_0$.  The output of the system is given
by $v(x)=|x|^n$, with $n$ a positive constant.  The SNR is defined,
as usual, by
\begin{equation} \label{SNR}
\mbox{SNR}=S(\omega_0)_{v(x)}/N(\omega_0)_{v(x)} \;\; ,
\end{equation}
where $S(\omega_0)_{v(x)}$ and $N(\omega_0)_{v(x)}$ are the output
signal and output noise corresponding to $v(x)$, respectively.

Under the transformation $\tilde x=e^\gamma x$ and $\tilde
D=e^{2\gamma} D$, with $\gamma$ constant, Eqs.  (\ref{model}) and
(\ref{SNR}) remain unchanged if
\begin{equation} \label{master}
f(x,t)=f(xe^\gamma,t) \;\; .
\end{equation}
Consequently, for the class of systems for which the previous
equality holds, for a certain value of $\gamma$, the SNR has the
same value at D and at $e^{2\gamma}D$.  This fact occurs when
$f(x,t)=q(\ln(x),t)$, where $q$ is a periodic function of its first
argument, with periodicity $\gamma$ if $\gamma$ is the lower
positive number satisfying Eq.  (\ref{master}).  Therefore, the SNR
is a periodic function of the logarithm of the noise level.  Notice
that both the signal and noise are not invariant under this
transformation, but they are changed in the following fashion
\begin{eqnarray}
S(\omega_0)_{v(\tilde x)}&=&e^{2\gamma n}S(\omega_0)_{v(x)}
\;\;, \nonumber \\
N(\omega_0)_{v(\tilde x)}&=&e^{2\gamma n}N(\omega_0)_{v(x)}
\label{SandN}
\;\;.
\end{eqnarray}

\begin{figure}[th]
\centerline{
\epsfxsize=6cm 
\epsffile{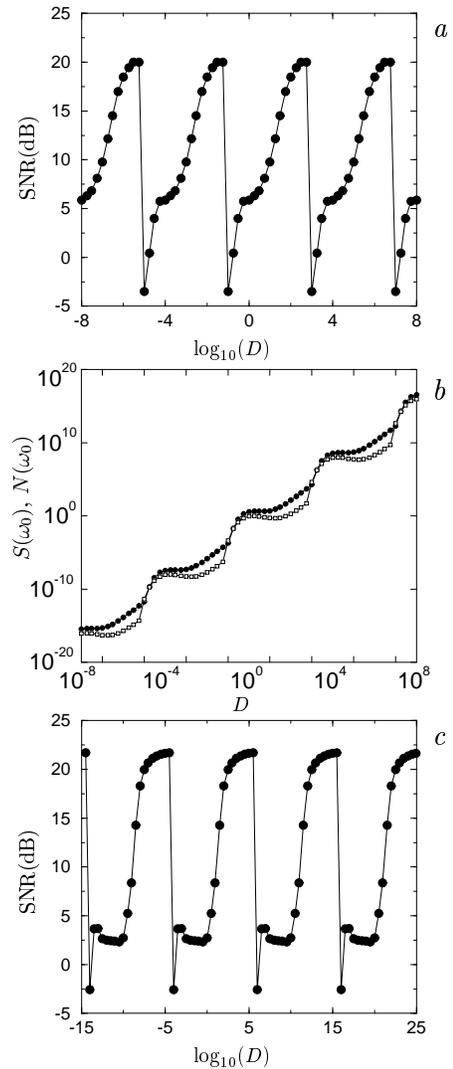}}
\caption[b]{\label{fig1}
(a) SNR corresponding to $f(x,t)$ given through Eq.  (\ref{ej1})
with parameter values $k_1=100$, $k_2=1$, $T=4$, $\beta=1$,
$\alpha=0.5$ and $\omega_0/2\pi=1$.  (b) Signal (filled circles) and
noise (empty squares) for the previous situation.  (c) SNR as in
Fig.  (a) but with $T=10$.
}
\end{figure}

In order to illustrate the previous results we have analyzed some
representative explicit expressions of $f(x,t)$.  To this purpose we
have numerically integrated the corresponding Langevin equation by
means of a standard second-order Runge-Kutta method for stochastic
differential equations \cite{kloeden}.  Moreover, as the output of
the system we have used $v(x)=x^2$.  We will first consider the case
in which
\begin{equation} \label{ej1}
f(x,t)=\Theta_T(\log_{10}(x^2))[\beta+\alpha\cos(\omega_0 t)]
\end{equation}
where $\alpha$ and $\omega_0$ are constants, and $\Theta_T(s)$ is a
square wave of period $T$ defined by
\begin{equation}
\Theta_T(s)=\left\{
\begin{array}{lcl}
k_1 &  & \mbox{if} \; \sin(2\pi  s /T) > 0 \;\; , \\
k_2 &  & \mbox{if} \; \sin(2\pi s /T) \le 0 \;\; ,
\end{array}
\right.
\end{equation}
with $k_1$ and $k_2$ constants.  In Fig.  \ref{fig1}(a) we have
plotted the SNR corresponding to the previous form of $f(x,t)$, for
particular values of the parameters.  This figure clearly manifests
the periodicity of the SNR as a function of the noise level and the
presence of multiple maxima at $D=D_0e^{mT}$, with $m$ being any
integer number and $D_0$ the noise level corresponding to the
maximum with $m=0$.  We show both the signal and noise in Fig.
\ref{fig1}(b).  This figure also corroborates the dependence of the
signal and noise on $D$ given in Eq.  (\ref{SandN}).

From this example one can infer the mechanism responsible for the
appearance of this phenomenon.  Due to the fact that the SNR has
dimensions of the inverse of time \cite{mio}, its behavior is
closely related to the characteristic temporal scales of the system.
Thus variations of the relaxation time manifest in the SNR.  In this
example, when $T$ is sufficiently large, for some values of the
noise level the system may be approximated by
\begin{equation}\label{ourmodel}
{dx \over dt} = -k_i[\beta+\alpha\sin(\omega_0 t)]x +
\sqrt{2D}\xi(t) \;\; ,
\end{equation}
where $i=1,2$, depending on the noise level.
In such a situation the SNR is given by
\begin{equation}\label{SNRex}
\mbox{SNR} = g(\alpha,\omega_0k_i^{-1})k_i \;\; ,
\end{equation}
with $g$ a dimensionless function \cite{mio}.  For a sufficiently
low frequency, the SNR is proportional to $k_i$ ($\mbox{SNR} \approx
g(\alpha,0)k_i$), i.e.  proportional to the inverse of the
relaxation time.  Consequently, there are two set of values of $D$
for which the SNR differs in approximately $10\log_{10}(k_1/k_2)$
dB, as one can see in Fig.  \ref{fig1}(c).  We then conclude that
multiple maxima in the SNR appears as a consequence of the form in
which the relaxation time of the system changes with the noise
level.

\begin{figure}[th]
\centerline{
\epsfxsize=6cm 
\epsffile{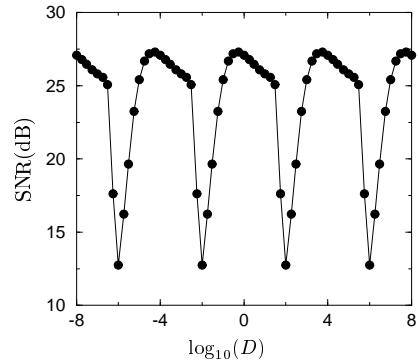}}
\caption[b]{\label{fig2}
SNR corresponding to $f(x,t)$ given through Eq.  (\ref{ej2}) 
parameter values $k_1=100$, $k_2=1$, with parameter values $k=100$,
$T=4$, $\beta=201$, $\alpha=100$ and $\omega_0/2\pi=1$.
}
\end{figure}

Let us now consider another explicit expression for $f(x,t)$ that
mainly differs from the previous one in the form in which the input
signal enters the equation,
\begin{equation} \label{ej2}
f(x,t)=k\sin(2\pi\log_{10}(x^2)/T)+[\beta+\alpha\cos(\omega_0 t)] \;\; .
\end{equation}
Here the spatial and temporal dependence of $f(x,t)$ appears in an
additive fashion instead of in a multiplicative way as in Eq.
(\ref{ej1}).  The results for the SNR are shown in Fig.
(\ref{fig2}) and also corroborate, in this case, the periodic
dependence on the logarithm of the noise level.

The importance of the class of systems discussed previously lies in
the fact that the periodicity of the SNR can be shown analytically
by simple considerations about the invariance of the system under
stretching transformations.  Numerical analyses have also revealed
that the presence of multiple maxima in the SNR as a function of the
noise level is a more general phenomenon than the situation
described by the equality (\ref{master}).  Thus, the appearance of
multiple maxima is robust upon variations of the form of $f(x,t)$.
For instance, if the periodicity in $s$ of the function $q(s,t)$ is
lost for high and low values of $s$ then the SNR may be a periodic
function of $\ln(D)$ for a bounded range of values of $D$.  In this
regard, we have performed numerical simulations and we have observed
that it is possible to obtain any number of maxima depending on the
form of $f(x,t)$, even if the periodicity of $q(s,t)$ does not hold
for any interval of $s$.  As an illustrative example we will analyze
a case in which the input signal enters $f(x,t)$ in a additive
fashion as well as in Eq.  (\ref{ej2}).

We will consider that the form of $f(x,t)$ consists of two
contributions; namely, one that comes from a time dependent
parabolic potential and the other without temporal dependence.
Then,
\begin{equation} \label{ej3}
f(x,t)=f_0(x)+[\beta+\alpha\cos(\omega t)] \;\; .
\end{equation}
Parabolic potentials may arise in many physical situations of
interest.  For instance, around a minimum most of the potentials may
be approximated by a parabolic one.  Experiments of a temporal
variation of the intensity of the potential are common, this is the
case of a single dipole under the influence of an external
oscillating field \cite{mio2}.  Another remarkable situation
described in the same way corresponds to some systems around a
bifurcation whose control parameter varies periodically in time, as
for example Rayleigh-B\'enard convection when the temperature
difference between plates varies slowly in a periodic fashion
\cite{ho,ga2}.  The term $f_0(x)$ may then represent a perturbation
to this ideal situation described by only a time-dependent harmonic
force.

\begin{figure}[th]
\centerline{
\epsfxsize=6cm 
\epsffile{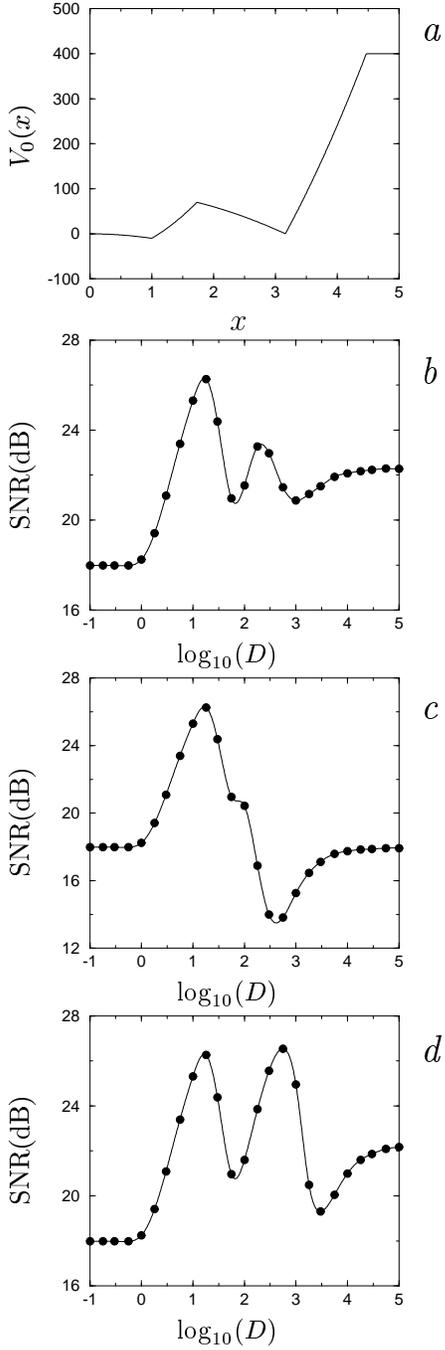}}
\caption[b]{\label{fig3}
(a) Potential $V_0(x)$ corresponding to the force $-f_0(x)x$ with
parameter values $k_1=20$, $k_2=-80$, $k_3=20$, $k_4=-80$, $k_5=0$,
$c_1=1$, $c_2=1.73$, $c_3=3.16$, and $c_4=4.47$.  (b) SNR
corresponding to $f(x,t)$ given through Eq.  (\ref{ej3}).  The
parameter values for $f_0(x)$ are the same as in case (a).  Moreover
$\beta=121$, $\alpha=100$, and $\omega_0/2\pi=1$.  (c) Same
situation as in case (b), except $k_5=20$.  (d) Same situation as in
case (b), except $c_4=7.75$.
}
\end{figure}

To be explicit and for the sake of simplicity, we will consider
$f_0(x)$ to be a linear piecewise function defined by
\begin{equation}
\label{perroverde}
f_0(x)=\left\{
\begin{array}{lcl}
k_1 &  & \mbox{if} \;\; |x| \le c_1 \;\; , \\
k_2 &  & \mbox{if} \;\; c_1 < |x| \le c_2 \;\; , \\
k_3 &  & \mbox{if} \;\; c_2 < |x| \le c_3 \;\; , \\
k_4 &  & \mbox{if} \;\; c_3 < |x| \le c_4 \;\; , \\
k_5 &  & \mbox{if} \;\; c_4 < |x|  \;\; . 
\end{array}
\right.
\end{equation}
In Fig.  \ref{fig3}(a) we have depicted the potential $V_0(x)$
corresponding to the force $-f_0(x)x$ for particular values of the
parameters $k_i$ and $c_i$.  It is worth emphasizing that we have
considered other forms of $f_0(x)$ but the same qualitative results
are obtained provided that $f_0(x)$ exhibits the main
characteristics of Eq.  (\ref{perroverde}), i.e.  the potentials
must have a similar location of maxima and minima.  In Fig.
\ref{fig3}(b) we have represented the SNR corresponding to the
previous function $f_0(x)$ as a function of the logarithm of the
noise level.  This figure clearly displays two maxima, one more
pronounced than the other.  Depending on the values of the
parameters the second maximum may disappear [Fig.  \ref{fig3}(c)] or
it can become more pronounced [Fig.  \ref{fig3}(d)].

In summary, we have found, for the first time, a class of systems
for which the response to a periodic force is enhanced, not only
with the addition of an optimized amount of noise, but also at
multiple values of the noise level.  Thus, the SNR exhibits a series
of maxima distributed periodically, as a function of the logarithm
of the noise level.  This feature has been found to be even more
general since any number of maxima may be present.  Among others, an
applied aspect to be emphasized concerns the possibility for the
design of devices for which the enhancement of an external signal
may occur at different values of the noise and not only at one
particular value.  Our findings, then, contribute to a wider
understanding of the phenomenon of SR by extending its scope and
perspectives, thereby embracing new situations that have not been
considered up to now.

This work was supported by DGICYT of the Spanish Government under
Grant No.  PB95-0881.  J.M.G.V.  wishes to thank Generalitat de
Catalunya for financial support.

\end{document}